\documentclass[preprint]{aastex}
\newcommand{\beq}{\begin{equation}}
\newcommand{\eeq}{\end{equation}}
\newcommand{\beqn}{\begin{eqnarray}}
\newcommand{\eeqn}{\end{eqnarray}}

\begin{document}

\title{On the Formation of Elliptical Rings in Disk Galaxies}

\author{Yu-Ting Wu$^1$ and Ing-Guey Jiang$^2$}

\affil{
{$^1$Institute of Astronomy and Astrophysics, 
Academia Sinica, Taipei, Taiwan}\\
{$^2$Department of Physics and Institute of Astronomy,}\\ 
{National Tsing-Hua University, Hsin-Chu, Taiwan} 
}

\begin{abstract}
N-body simulations of galactic collisions are employed to investigate 
the formation of elliptical rings in disk galaxies.
The relative inclination between disk and dwarf galaxies
is studied with a fine step of five degrees. 
It is confirmed that the eccentricity of elliptical ring
is linearly proportional to the inclination angle.
Deriving from the simulational results, an analytic formula 
which expresses the eccentricity 
as a function of time and inclination angle is obtained.  
This formula shall be useful for the interpretations
of the observations of ring systems, and 
therefore reveals the merging histories
of galaxies.

\end{abstract}


\section{Introduction}

Ring galaxies are galaxies which present a very unusual morphology in that rings are observed as dominating 
parts of their structures.
They could be systems with rings surrounding early-type galaxies,
such as the Hoag's object or polar ring galaxies (Hoag 1950; Wakamatsu 1990).
They could also be those systems
in that 
the main disk of spiral galaxies are dominated 
by a ring centered on a nucleus. 
This paper focuses on the second type, 
which is as those presented in the catalog of Madore et al. (2009).

Although the secular processes
through the bars might be responsible for some of the 
ring structures in spiral galaxies,
the galactic interactions through collisions or merging
are the main picture to describe the formation of ring galaxies.
The head-on collisions were first proposed by 
Lynds \& Toomre (1976) to be the mechanism to produce 
ring galaxies. The analytic model of these kinds of collisions
employing the caustic wavefronts of stellar rings was developed 
in Struck-Marcell (1990) and  Struck-Marcell \& Lotan (1990). 
Later, Gerber et al. (1992) studied the formation of empty rings, which are
those systems without nuclei such as Arp 147. They produced C-shape 
structures and found that the remnant nucleus 
is offset from the center and out of the original plane in their simulations.
Moreover, Wong et al. (2006) suggested that a high-speed off-center collision 
can explain their observational results of NGC 922. 

Interestingly, Elmegreen \& Elmegreen (2006) showed that 
some ring galaxies with features of 
collisional rings have no obvious companions.   
It is thus unclear whether 
these ring galaxies could be explained 
by the collision scenario or not. 
Motivated by this controversial result, 
Wu \& Jiang (2012) picked up some 
observed axisymmetric ring galaxies from 
the catalog of Madore et al. (2009) and demonstrated through
N-body simulations that
head-on collisions could produce the structures 
of these axisymmetric rings. 
They thus claimed that the identified off-center companions 
around the ring galaxies might not be relevant to the
rings. The dwarf galaxy which collided with the disk galaxy along the 
symmetric axis was already mixed with the center of the disk galaxy and 
became parts of a nucleus of this ring galaxy.

While these results were presented in IAU Symposium
(Wu \& Jiang 2011), in addition to our paper Wu \& Jiang (2012),
there were several recent works, such as Smith et al. (2012),
Mapelli \& Mayer (2012),  and
Fiacconi et al. (2012),  
which also investigated the formation of ring galaxies.
Smith et al. (2012) modeled the formation of Auriga's Wheel ring
galaxy using N-body simulations.
They proposed that Auriga's Wheel
was formed by the collision of a spiral galaxy and an elliptical
galaxy.
Mapelli \& Mayer (2012) investigated the formation
of empty ring galaxies, which were considered as collisional ring galaxies
without nuclei inside the ring.
Fiacconi et al. (2012) explored parameters of 
galactic collisions and paid attention to the
relation between the star formation rate and different parameters.
However, 
the association between different parameters and the
shape of the ring galaxies was not clear.

Indeed, as those observed non-axisymmetric rings shown in  Madore et al. (2009), 
it is confirmed that not only  
the shape of rings could be elliptical, but also that the nucleus 
could be offset from the center of rings. 
These complexities are likely due to the details of 
the collisional history. 
The interesting questions would be: what kinds of 
relative orbits between the dwarf and main disk galaxies could lead 
to these structures?  

In order to answer the above question, 
the link between the merging processes and the shapes of rings  
needs to be constructed.
Once this link is established, through the observations of ring systems,
one can investigate the merging histories of galaxies 
and therefore constrain the cosmology. 

To achieve this goal, we here
employ N-body simulations to investigate the 
formation of ring galaxies through non-axisymmetric collisions.
The collisions of two galaxies would be set up
by making the initial velocity of the 
dwarf be inclined from the symmetric axis of the main disk galaxy.  
To narrow down the parameter space and focus on the
inclination angle, the dwarf galaxy is assumed to be moving 
on a parabolic orbit and directly towards to the center of the disk galaxy.

Moreover, in order to describe the resulting ring,
we develop a new procedure
to determine the eccentricities of rings 
and the nucleus offset.  
With these quantitative description, we will be able to 
study the relation between the shapes of rings 
and the inclinations of galactic collisions.

In the following sections,  
the model is given in \S 2, and the descriptions and results
of simulations are in \S 3. We provide conclusions in \S 4.  

\section{The Model}

A disk galaxy and a dwarf galaxy are set as a target and an intruder 
in the simulations to investigate the outcome of collisions 
between two galaxies.
The disk galaxy is composed of three components: 
a stellar bulge, a stellar disk, and a dark halo. 
The dwarf galaxy is consisted of a stellar component and a dark halo.

\subsection{The Profiles}

The density profiles of the disk galaxy, including the stellar bulge, 
the stellar disk, 
and the dark halo, follow the models in Hernquist (1990, 1993). 
For the stellar bulge, the density 
follows the Hernquist profile (Hernquist 1990), and is given by
\begin{equation}
 \rho_{b}(r)=\frac{M_b}{2 \pi} \frac{b}{r} \frac{1}{(r+b)^3},
 \label {eq2-1}
\end{equation}
where $M_b$ is the bulge mass and $b$ is the scale length.
Besides, the density profile of the stellar disk is 
\begin{equation}
 \rho_{d}(R,z)=\frac{M_d}{4 \pi h^2 z_0} \exp(-R/h) {\rm sech}^2(\frac{z}{z_0}),
 \label {eq2-2}
\end{equation}
where $M_d$ is the disc mass, ${\it h} $ is the radial scale length and ${\it z_0}$ is the vertical scale length.
The dark halo's density profile is
\begin{equation}
 \rho_{h}(r)= \frac{M_h}{2 \pi ^{3/2}}\frac{\alpha}{r_t r^2_c}\frac{\exp(-r^2 / r^2 _t)}{\frac{r^2}{r^2_c}+1} ,
 \label {eq2-3}
 \end{equation}
where $M_h$ is the halo mass, $r_t$ is the tidal radius, and $r_c$ is the core radius.
The normalization constant $\alpha$ is defined as
 \begin{equation}
 \alpha = \left \{ 1-\sqrt{\pi}q \exp(q^2)[1-{\rm erf}(q)]\right \}^{-1},
 \label {eq3-3}
 \end{equation}
where $q = r_c / r_t$ and ${\rm erf}(q)$ is the error function as a function of $q$.

The density profile of the intruder dwarf galaxy, which is comprised of 
a dark halo and a stellar part, follows the Plummer sphere 
(Binney \& Tremaine 1987; Read et. al. 2006): 
 \begin{equation}
 \rho_{\rm plum}(r)= 
 \frac{3M_p}{4 \pi a^{3}}\frac{1}{\left ( 1+ \frac{r^2}{a^2} \right )^{2/5}},
 \label {eq3-4}
 \end{equation}
where $M_p$ and $a$ are the mass and the scale length of the halo 
or stellar component.

The initial positions of particles are given according to the 
density profiles 
as described above. 
In addition, the initial velocities of particles in spherical systems can 
be determined 
from phase-space distribution which is calculated from Eddington's formula 
(Binney, Jiang \& Dutta 1998).
On the other hand, for a non-spherical system, such as the stellar disk, 
the particles' velocities are determined from the moments of 
the collisionless Boltzmann equation as described in Hernquist (1993).

\subsection{The Parameters}
The parameters of the disk galaxy and the dwarf galaxy 
are summarized in Table 1 and Table 2, respectively.
All parameters of the disk galaxy 
are the same as those in Wu \& Jiang (2012), 
except that the number of particles is increased five times 
and the stellar bulge is added. 
The total masses of the stellar disk and dark halo are exactly
the same as the values used in Hernquist (1993),
which are from the mass model of a Sb type spiral galaxy.
The parameter of the dwarf galaxy is also the same as 
the dwarf galaxy $\rm DG_A$ in Wu \& Jiang (2012) except the increment of the particle number.

\begin{table}[h]   
\begin{center}
\caption[Model parameters of the disk galaxy]{
Model parameters of the disk galaxy.
}
 \begin{tabular}{lcccc}
 \hline
 \hline
 \multicolumn{2}{l}{Stellar Bulge}&&&  \\
 \hline
 mass $M_b$ ($10^{10} M_{\odot}$) & 1.68 && \\
 scale length $b$ (kpc) & 0.7 &&\\
 number of particles & 75,000 &&\\
 \hline
 \multicolumn{2}{l}{Dark Halo}&&& \\
 \hline
 mass $M_h$ ($10^{10} M_{\odot}$) & 32.48 &&\\
 core radius $r_c$ (kpc) & 3.5 &&\\
 tidal radius $r_t$ (kpc) & 35.0 &&\\
 number of particles & 1,450,000 &&\\
 \hline
 \multicolumn{2}{l}{Stellar Disc}&&&  \\
 \hline
 mass $M_d$ ($10^{10} M_{\odot}$) & 5.6 && \\
 radial scale length $h$ (kpc) & 3.5 &&\\
 vertical scale-height $z_0$ (kpc) & 0.7 &&\\
 number of particles & 250,000 &&\\
 \hline
 \end{tabular}
\label{table:disk}
\end{center}
\end{table}

\begin{table}[h]  
\begin{center}
\caption[Model parameters of the dwarf galaxy]{
 Model parameters of the dwarf galaxy.}
 \begin{tabular}{lcc} 
 \hline
       & \multicolumn{2}{c}{Dwarf Galaxy}\\
 \cline{2-3} 
 & Dark Halo & Stellar Part\\
 \hline
 mass $M_p$ ($10^{10} M_{\odot}$) & 7.616 &1.904\\ 
 scale length $a$ (kpc) & 3.0 &1.5\\
 number of particles & 340,000 & 85,000\\
 \hline
 \end{tabular}
 \label{table:dwarf}
\end{center}
\end{table}

The simulations are performed with the parallel tree-code
GADGET (Springel et al. 2001). 
The unit of length is 1 kpc, the unit of mass is $10^{10} M_\odot$, 
the unit of time is $9.8\times10^8$ years, and  
the gravitational constant $G$ is 43007.1.
 
Using the parameters and the unit above, the dynamical time of 
the disk galaxy, $T_{dyn}$, can be calculated 
by $T_{dyn} \equiv 2 \pi R_{1/2}/ v_{1/2}=0.157$, 
where $R_{1/2}=5.88$ is the disk’s half-mass radius 
and $v_{1/2}=235.82$ is the velocity of a test particle 
at $R_{1/2}$.

\subsection{The Initial Equilibrium}

The stellar component and the dark halo
of the dwarf galaxy are both spherical, so
can be set up simultaneously 
in equilibrium initially.
On the other hand, the spherical components 
(the stellar bulge and the dark halo) and 
non-spherical component (the stellar disk) of the disk galaxy are 
set up separately, 
and are combined together to be the disk galaxy model. 
Then, the disk galaxy is allowed to relax towards 
a new equilibrium state.

Using the same method in Wu \& Jiang (2009), the virial theorem 
is used to check 
whether the disk galaxy is already in a new equilibrium. 
Based on the virial theorem, when the disk galaxy is in equilibrium, 
the value of virial ratio, $2K/|U|$, should be around 1, 
where $K$ and $U$ are the total kinetic energy and the total 
potential energy, respectively.

Fig. 1 shows the virial ratio $2K/|U|$ as a function of time in the top panel, 
and the Lagrangian radii as a function of time in the bottom panel.
In the bottom panel, the curves from bottom to top show the radius enclosing 
$10\%, 25\%, 50\%, 75\%, 90\%$ of the total mass. 
From this figure, it can be seen that the disk galaxy approaches a new equilibrium at t = 15$T_{dyn}$.
Besides, the energy is conserved as there is only $0.01\%$ energy variation.
Therefore, the disk galaxy at $t=15 T_{dyn}$ of relaxation 
is used as the target disk galaxy at the beginning of 
collision simulations.

\section{The Descriptions and Results of Simulations}

The initial separation of the disk galaxy and the dwarf galaxy is set 
to be 200 kpc to make sure that two galaxies are well separated initially.
The initial relative velocity, $v_i=143.1$ km/s, is given according to the 
orbital energy and 
the parabolic orbit is considered here.

To investigate the effects of the inclination angle 
on the formation and evolution of asymmetric rings, 
13 simulations with inclination angle $i$ ranging from 0 to 60 degrees
are presented in this paper, 
and named Si00, Si05, Si10, Si15, and so on, until Si60.
The definition of the inclination angle $i$ is the angle between 
the symmetry axis of the disk galaxy and 
the line connected between the center of mass of the disk galaxy and the center of mass of the dwarf galaxy.
The number, such as 00, 05, 10 etc., after the characters $Si$ represents 
different inclination angle, such as 0, 5 and 10 degrees. 
Thus, in simulation Si00, 
the disk galaxy and the dwarf galaxy are both moving along $z$ axis.
In all other simulations, two galaxies' orbits are on 
the $y-z$ plane of the coordinate system.

The reason why the maximum inclination angle $i=60$
is that it is difficult to obtain a complete ring for even larger $i$.
The increment of the inclination angle between successive simulations is
chosen to be five degrees, which is sufficient for our study and 
presents a much higher resolution than previous work. 
Besides, in order to understand detail processes during each simulation, 
the time interval between each snapshot 
is one-tenth of the dynamical time 
and defined as $T_s\equiv 0.1 T_{dyn}$.

\subsection{The Collision}

The dynamical evolution of both disk and dwarf galaxies
are presented in this subsection.
The encounter processes
in three simulations Si00, Si20, and Si40 are shown in Fig. 2, 
and the evolution of
stellar density distribution of Si40 
is summarized in Fig. 3-4.
Though only three simulations, out of 13, are shown here,  
the principle dynamical processes are carefully investigated.
  
Fig. \ref{fig3:5sim_cmcv} shows the centers of mass of the disk 
and the dwarf galaxies on the $y$ and $z$ axis as a function of time 
in simulations Si00, Si20, and Si40.
The solid and long dashed curves show the centers of mass of 
the whole disk galaxy 
and the whole dwarf galaxy, respectively.
 The short dashed and the dotted curves correspond to the centers of mass 
 of the disk's stellar component (including the bulge and the disk parts) 
 and the dwarf's stellar component.
 
 For Si00, i.e. Figs. \ref{fig3:5sim_cmcv} (a) and (b), 
 the center of mass on the $y$ axis is around 0 
 because the inclination angle $i=0$ at the beginning of the simulation.
 Besides, the disk galaxy and the dwarf galaxy  
 collide with each other at $t=59T_s$ and then well separate at $t=70T_s$.
 At $t=80T_s$, they become close again.
 Due to the dynamical friction, the dwarf galaxy loses some
kinetic energy and two galaxies' cores are merged.
However, after the first close encounter, 
because many outskirt particles of the main galaxy continue 
to move upward, i.e. $+z$ direction,
and many particles of the dwarf galaxy move toward  
$-z$ direction, the center of mass of the main galaxy 
always has a positive $z$ coordinate and 
the center of mass of the dwarf galaxy
always has a negative $z$ coordinate even after the second 
encounter. 

For Si20 (Figs. \ref{fig3:5sim_cmcv} (c) and (d)) 
 and Si40 (Figs. \ref{fig3:5sim_cmcv} (e) and (f)),
 because the inclination angles are 20 and 40 degrees 
 at the beginning of simulations, 
 the centers of mass on the $y$ and $z$ axis vary as a function of time.
 Two galaxies also have the first close encounter at $t=59T_s$, 
 and the second encounter around $t=79T_s$.
 It shows that the initial inclination angle 
 in simulations does not affect the encounter time very much.
 
 Another interesting feature in Fig. \ref{fig3:5sim_cmcv} is that 
 the center of mass of the whole dwarf galaxy is very different from 
 the center of mass of the dwarf's stellar component after the time 
 at which two galaxies are separated at their furthest distance 
 after their first encounter, 
 i.e., $t=70T_s$.
 This is because the outer part of the dwarf's halo
gets destroyed, becomes  
non-symmetric, and escapes far away from the main part 
of the dwarf galaxy after the collision.

On the other hand, 
Fig. \ref{fig3:i00s4_xy_2D} and Fig. \ref{fig3:i00s4_yz_2D} 
give an example of 
the time evolution of the projected density of all stellar particles, 
 including the stellar disk, the bulge, and the dwarf's stellar component, 
 on the $x-y$ and $y-z$ plane in the simulation Si40. 
At the beginning, i.e. $t=0$, the center of mass of the dwarf galaxy 
and the disk galaxy 
are separated with a distance 128.5 kpc on the y-axis 
and 153.2 kpc on the z-axis.
Then, they approach and pass through each other 
from $t=58T_s$ to $t=59T_s$. 
Being a collision with inclination angle $i=40$ degrees, 
an elliptical ring is formed at $t=60T_s$.
The ring is not azimuthally uniform, 
and the center of the disk galaxy shifts towards the $+y$ direction.
Later on, the ring expands and the density of the ring decreases, 
as shown at $t=65T_s$. 
Due to the projection effect, it seems that, on the $x-y$ plane,
the dwarf galaxy is connected with the ring.  
Besides, the location of the disk center, which has the highest projected 
density, 
does not move a lot, but the ring expands more towards the $+y$ direction.
 During this stage, the dwarf galaxy keeps moving away from the disk galaxy.
 From $t=70T_s$ to $t=75T_s$, 
the dwarf galaxy returns and has the second encounter. 
 
Moreover,
Fig. \ref{fig3:com_xy_scatter} and Fig. \ref{fig3:com_yz_scatter}
are those snapshots 
in Fig. \ref{fig3:i00s4_xy_2D} and Fig. \ref{fig3:i00s4_yz_2D} but are
plotted in terms of particle positions. 
The black dots represent the particles of the stellar disk and the bulge 
of the disk galaxy, 
while the red dots represent the dwarf's stellar particles. 
The evolution of particle distributions of both galaxies
can therefore be seen clearly. 
After the first collision, some dwarf's stellar particles 
get trapped  
near the center of the disk galaxy,
some have gone very far from the cores of both galaxies, and 
a certain fraction of particles follow the orbit of the dwarf
galaxy.  Comparing Fig. 5 with Fig. 3, 
it is obvious that the ring is mainly made of particles originally belonging
to the disk galaxy.
 In the end, at $t=82T_s$, the disk galaxy and the dwarf galaxy merge.

In general, when a satellite galaxy falls into a major galaxy, 
the satellite experiences
the dynamical friction from the stars and halo of the major galaxy.
The satellite would lose some energy and have an orbital decay. 
A similar process was shown in Jiang and Binney (2000).

\subsection{The Ring}

From the snapshots in Figs. 3-4., it is likely that the collisions 
 with non-zero inclination angle $i$ could produce elliptical rings. 
In order to understand the shape of elliptical rings,
we develop a new method to determine the related parameters, 
such as semi-major axis, semi-minor axis, and eccentricity.
The procedure is described below 
with an example demonstrated in Fig. \ref{fig3:fitring}.

First, the stellar components of the disk galaxy, including the bulge 
and the stellar disk, 
are projected on the $x-y$ plane. 
Then, we divide the space into small square cells, 
 and calculate the projected density of the particles within each cell.
 The size of each cell is 0.2 kpc $\times$ 0.2 kpc. 
After that, taking the center of the cell which has the highest 
projected density as the coordinate center, 
we divide the space equally into 36 cells in the azimuthal direction 
as shown in Fig. \ref{fig3:fitring} (a).
Fig. \ref{fig3:fitring} (a) shows the projected density 
on the $x-y$ plane at $t=60T_s$ in simulation Si30.
The black lines represent the grid-lines, and the angle between 
two grid-lines is ten degrees.
Following this, we calculate the accumulated number of particles $N_{acc}$ 
contained in each azimuthal
cell as a function of radius $R$ 
(with 200 particles in each radial bin), 
as shown in Fig. \ref{fig3:fitring} (b).
 From Fig. \ref{fig3:fitring} (b), it shows that $N_{acc}$ increases 
rapidly within 0.5 kpc and 
then slows down from 0.5 to 1.5 kpc. 
After $R=$ 1.5 kpc, $N_{acc}$ increases rapidly again until R$\sim$4 kpc. 
Obviously, the concentration in the galactic center 
produces the rapid increment of $N_{acc}$ within 0.5 kpc, 
and the existence of a ring explains another rapid 
increment between 1.5 and 4 kpc.  
 
To identify the ring region precisely,
the slope of the curve in Fig. \ref{fig3:fitring} (b) 
is plotted as a function of $R$ as shown in Fig. \ref{fig3:fitring} (c). 
Thus, the black dots are the slope $m$ at different radius $R$. 
The red triangle labels the maximum of the slope, $m_{max}$, 
which should be located within the central part of the ring region.
 Between the disk center and the ring region, the blue triangle 
labels the minimum of the slope, $m_{min}$. 
 Thus, the average of the maximum slope and the minimum slope could be 
calculated as $m_{ave}=(m_{max}+m_{min})/2$ and 
the corresponding inner radius $ri$ and outer 
radius $ro$, where the slopes are  equal 
to $m_{ave}$, could be determined, 
as represented by the green dots in panel (c).
The above is done for all 36 azimuthal cells.
Therefore, for each azimuthal cell, 
we define the inner boundary of ring region to be 
$ri$,
the outer boundary of ring region to be $ro$,
the radial coordinate of the ring to be $(ri+ro)/2$. 
Thirty-six black dots in Fig. \ref{fig3:fitring} (d) 
indicate the radial coordinates of the ring at  
different azimuthal cells, and 
the green dots represent the locations of $ri$ and $ro$.

Finally, those black dots are used to determine an ellipse
following Gander et al. (1994), in which 
Gauss-Newton method (Gill, Murray \& Wright 1981) is used.
The best-fitting ellipse is obtained 
when the sum of the squares of the distances 
from the black dots to the ellipse is minimized.
Thus, the black ellipse in Fig. \ref{fig3:fitring} (d) 
shows our best fitting ellipse. 
The semi-major axis $R_a$, the semi-minor axis $R_b$, 
the eccentricity $e$, and the center  
of this ellipse can be determined correspondingly.
 
Moreover, because   
the elliptical ring is a pattern generated during the galactic collision, 
its center could be offset from the disk center.
In order to determine this offset,
we draw a straight line starting from the ellipse center, passing the 
disk center, and arriving at the ellipse boundary.
The total length of this line $l$ and also   
the distance $d_{ed}$ from the ellipse center to the disk center
can be calculated. 
The offset $p$, which is a dimensionless parameter,
is defined as
 \begin{equation}
 p=\frac{d_{ed}}{l}.
 \label {eq4-p}
 \end{equation}
Thus, when the disk center is close to the ellipse center,
the offset $p$ is close to 0;
when the disk center is close to the boundary of the ellipse, 
the offset $p$ is close to 1. 
 
Here we present the results determined through the above method.
Fig. \ref{fig3:com_an} (a) shows the time evolution of 
the semi-major axis of rings in all 13 simulations. 
The lines with different colors and different symbols 
correspond to the simulations with different inclination angles, 
as indicated in panel (a).
The rings all form at $t=60T_s$ but exist in a shorter duration when 
the collisions have larger inclination angles. 
 For example, the lifetime of the ring in Si60 is only 2$T_s$, 
that is from $t$=60 to $t$=62$T_s$.
To understand the effect of the inclination angle 
on the semi-major axis of the ring, 
the ring in Si00 is used as a standard. 
Fig. \ref{fig3:com_an} (b) shows the semi-major axis 
of the ring as a function of time in Si00 only.
Then, Fig. \ref{fig3:com_an} (c) shows the differences of 
the semi-major axis of the ring, $\Delta R_a$, between 
other simulations and Si00 at different times.
For example, the semi-major axis of the ring at $t$=62$T_s$ in Si00 
and Si40 are 7.71 and 8.15, respectively.
Therefore, $\Delta R_a$ between the simulation Si00 and Si40 at $t$=62$T_s$ 
is 0.44, as shown in panel (c) 
 (red line with dotted symbols).
Thus, from panel (c), we can see that there is no significant 
difference between
Si00 and the simulations with inclination angles 
smaller than 15 degrees.
It means that the rings produced by the collisions with 
inclination angles smaller than 15 degrees are similar 
as the rings produced by the head-on collision.

Finally, Fig. \ref{fig3:com_an} (d) presents the differences 
between the semi-major axis, $R_a$, and 
 the semi-minor axis, $R_b$, of these rings as functions of time. 
The difference between these two axes, 
i.e. $R_a-R_b$, is larger 
when the collision has larger inclination angle.
Please note for the head-on collision, i.e. Si00,
$R_a-R_b$ is very small but not completely zero. 
This leads to an eccentricity about 0.1. 
Thus, in this paper, any ring with 
eccentricity about 0.1 is considered to be 
equivalent to a circular ring.
In fact, as we can see in panel (c), all simulations with 
inclination angle smaller than 15 degrees
present very similar results and to distinguish them is 
beyond the scope of this paper.   
 
On the other hand, because $R_a$ increases with time, 
the out-moving speed $v_{om}$ of the semi-major axis, $R_a$, 
is defined as $(R_a(T_2)-R_a(T_1))/T_s$, 
where $R_a(T_2)-R_a(T_1)$ is the variation of the semi-major axis 
within one $T_s$ from $T_1$ to $T_2$. 
We also define the out-moving speed 
of the semi-minor axis similarly.
The average of the out-moving speed for all simulations is shown 
in  Fig. \ref{fig3:com_vel}.
The black line presents the out-moving speed of 
the semi-major-axis, and the red line is for the semi-minor axis.
The error bar indicates one standard deviation.
The largest standard deviation in the figure is 6.1 km/s.
Thus, the inclination angle has no significant effect 
on the out-moving speeds of ring patterns. 
In addition, it is apparent that the out-moving speeds of 
the semi-major axis and the semi-minor axis 
are similar and both decrease gradually as functions of time.

The results of the eccentricity and offset of rings are presented here.
The eccentricity can be calculated easily from 
the semi-major axis $R_a$ and semi-minor axis $R_b$.  
Fig. \ref{fig3:com_an_2} (a) presents the time evolution 
of the ring eccentricity $e$ in all 13 simulations.
Different lines represent different simulations as indicated 
in Fig. \ref{fig3:com_an} (a).
When the inclination angle is smaller than 15 degrees,
the ring eccentricity is about 0.1. 
On the other hand, for the simulations with inclination angles larger 
than 15 degrees, the eccentricity of the ring decreases slowly with time.
  
Fig. \ref{fig3:com_an_2} (b) presents the offset $p$ as a function of time.
Different lines are for different simulations as in panel (a).
It shows that the disk center is very close to the center of the ellipse 
when the inclination angle of the collision is 0, i.e. the simulation Si00. 
Besides, for all simulations, the offset $p$ is larger 
at the beginning and then becomes smaller at $t=61 T_s$.
This is because at the beginning of the collision, the disk center  
moves towards $+y$ direction 
and gets closer to the boundary of the ellipse, 
as the example shown in Fig. \ref{fig3:i00s4_xy_2D} $t=60T_s$. 
After that, the ring expands, 
and the ellipse center shifts toward $+y$ direction as well. 
However, the location of the disk center does not move significantly 
after $t=60T_s$.
Thus, at $t=61T_s$, the disk center and the ellipse center 
are close to each other and the offset $p$ becomes smaller.
After $t=61T_s$, the ellipse center shifts toward $+y$ direction 
continuously, so the offset $p$ increases slightly.
 
Fig. \ref{fig3:com_an_2} (c) and (d) present
the ring eccentricity $e$ and the offset $p$ as functions of 
the inclination angle of the collisions, respectively.
For these two panels, different lines represent different  
times, as indicated in panel (c).
From panel (c), we can see that for the collisions 
with inclination angle larger than 15 degrees, 
the eccentricity increases steadily.
 For example, at $t=60T_s$, the eccentricity in Si20, Si40 and Si60 
are 0.25, 0.48 and 0.71, respectively.
From panel (d), after $t=62T_s$, it can be seen that at the same 
simulation time, 
the ring has larger offset $p$ when the collision has larger 
inclination angle. 

In fact, the most striking feature shown in Fig. \ref{fig3:com_an_2} (c)
is that  for $i=15,20,...,55$ degrees, and $t=60,61,...,64 T_s$,
the results are simply straight lines with slightly different slopes. 
The information of slope differences is actually in  
Fig. \ref{fig3:com_an_2} (a), which shows that the eccentricity varies 
with time as $-\sqrt t  $ approximately.
Thus, using the data points for the above $i$ and $t$, 
an analytic formula expressing 
the eccentricity as a function of $t$ and $i$ on $t-i$ plane 
can be obtained: 
 \begin{equation}
 e(t,i)= \{ \beta_1 -\beta_2 
\sqrt{ t/T_s-t_f/T_s }\} {i}/{\rm degree}, 
\label {eq5-eti}
 \end{equation}
where $\beta_1 = 0.0120$, $\beta_2=0.0025$, 
$t_f=60 T_s$. 
Though this equation is
unlikely to fit all simulational results perfectly. 
It is a good approximation
which could be a useful tool for 
the interpretations of elliptical ring systems.

\section{Conclusions}

Motivated by the existence of elliptical rings in disk galaxies,
the effect of the inclination angle between the merging pair 
on the morphology of elliptical rings is investigated. 
This is the first time to have
such high resolution on the inclination
and thus the association between the 
inclination and the shape of ring galaxies 
are quantified. 

The results have shown that the elliptical rings could be produced 
by the collisions with inclination angles from 15 to 60 degrees.
The elliptical ring's eccentricity ranges from 0.1 to 0.7 and the offset $p$
is from nearly 0 to about 0.25.
The out-moving speeds of ring patterns are not 
affected by the inclination but decays with time,
from about 170 km/s  to  90 km/s.

We confirm that the galaxy pairs with larger inclination angles 
can produce elliptical rings with larger eccentricities.
The linear dependence between eccentricity and inclination 
leads to an analytic formula in which the eccentricity
is expressed as a function of time and inclination.
This simple rule deriving from the results of N-body simulations 
shall be very helpful for the future investigation
of merging histories of galactic systems with 
elliptical rings.  

\section*{Acknowledgment}
We are thankful for the referee's good suggestions.
We are grateful to Ronald Taam for the discussions.
We also thank the National Center for High-performance Computing
for computer time and facilities. 
This work is supported in part 
by the National Science Council, Taiwan, under 
NSC 100-2112-M-007-003-MY3.


\clearpage

\begin{figure}[h]
\begin{center}
\includegraphics[angle=0,scale=.60]{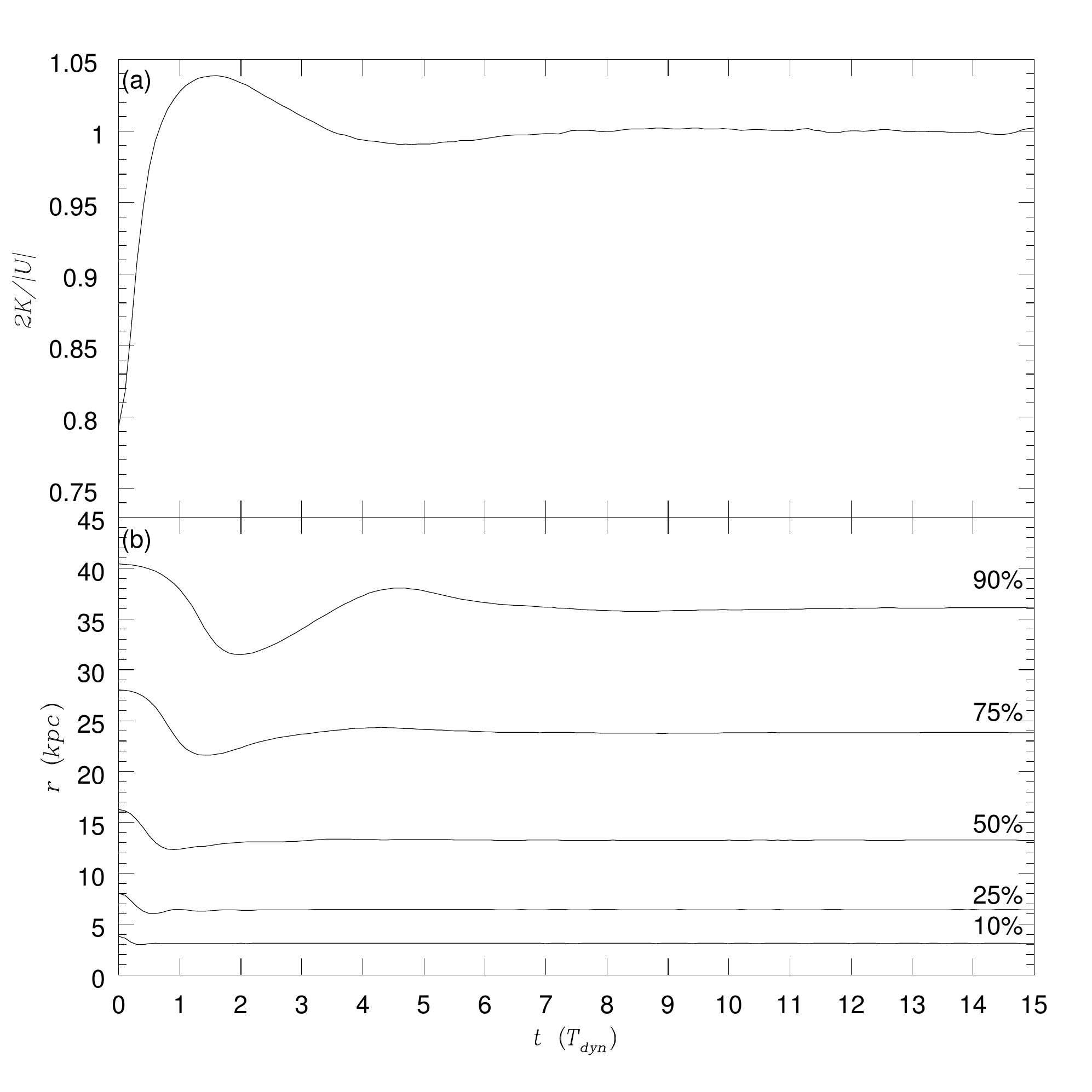}
\caption[The virial ratio $2K/|U|$ and Lagrangian radii of the disk galaxy as a function of time $t$.]{The virial ratio $2K/|U|$ (top panel) and 
Lagrangian radii (bottom panel) of the disk galaxy
as a function of time $t$, where the unit of $t$ is its dynamical time.} 
\label {fig:BHD_sM_5_s}
\end{center}
\end{figure}
\clearpage     
\begin{figure}
 \begin{center}
\includegraphics[trim=0cm 0.5cm 0cm 0cm,clip=true,angle=0,scale=0.9]
{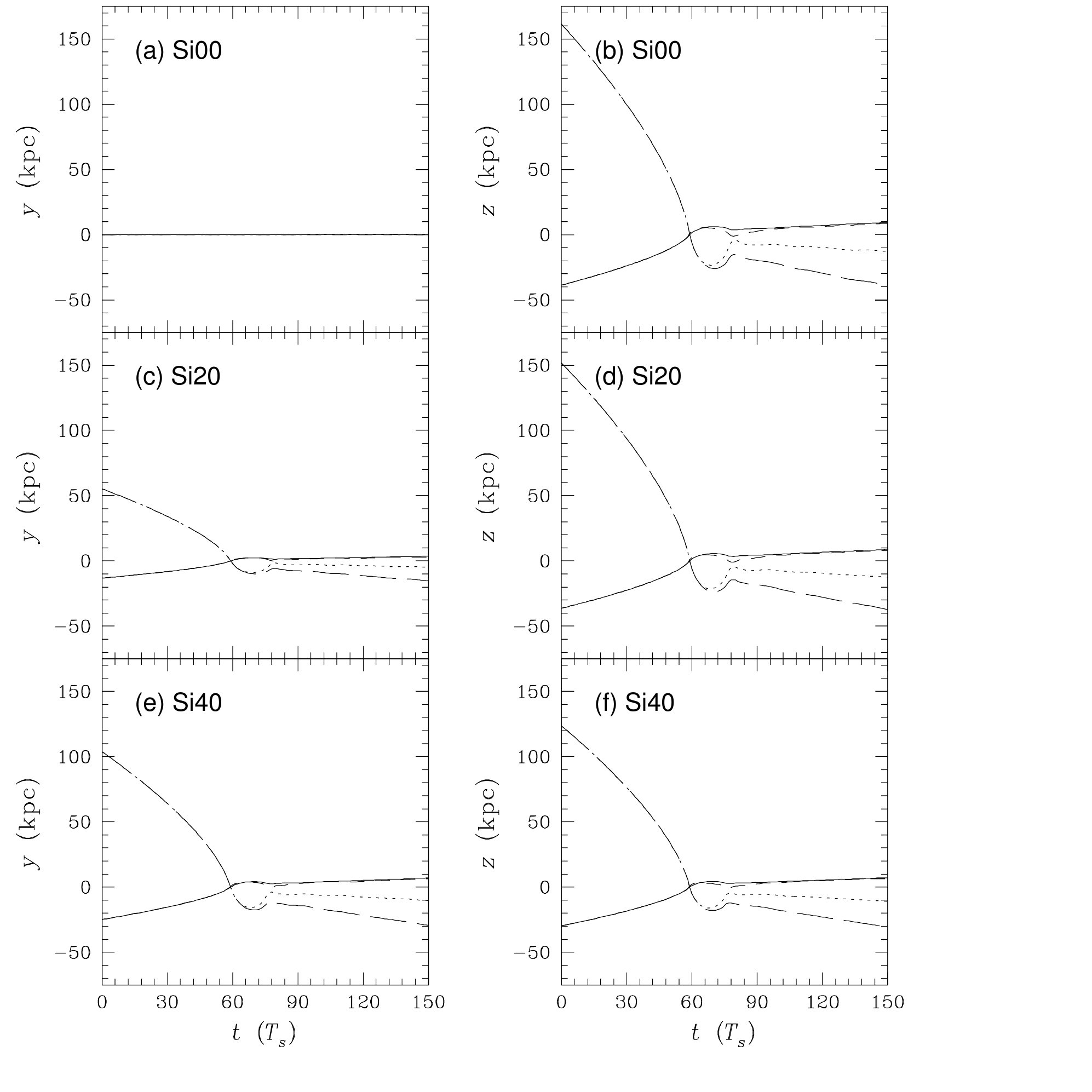}
\caption[The evolution of simulations Si00, Si20 and Si40.]
{The evolution of simulations Si00, Si20 and Si40.
 Panels (a)-(f) show the locations of the disk galaxy 
and the dwarf galaxy on $y$ and $z$ axis as  
functions of time in Si00, Si20 and Si40, respectively.
 The solid and long dashed curves correspond to the centers of mass of 
 the whole disk galaxy and the whole dwarf galaxy.
 The short dashed and dotted curves represent the centers of mass of 
 the disk's stellar components (including both the bulge and the stellar disk) 
 and the dwarf's stellar component.}
 \label {fig3:5sim_cmcv}
 \end{center}
\end{figure} 
\clearpage
\begin{figure}
 \begin{center}
 \includegraphics[trim=0cm 1.8cm 0cm 1.5cm,clip=true,angle=0,scale=0.7]{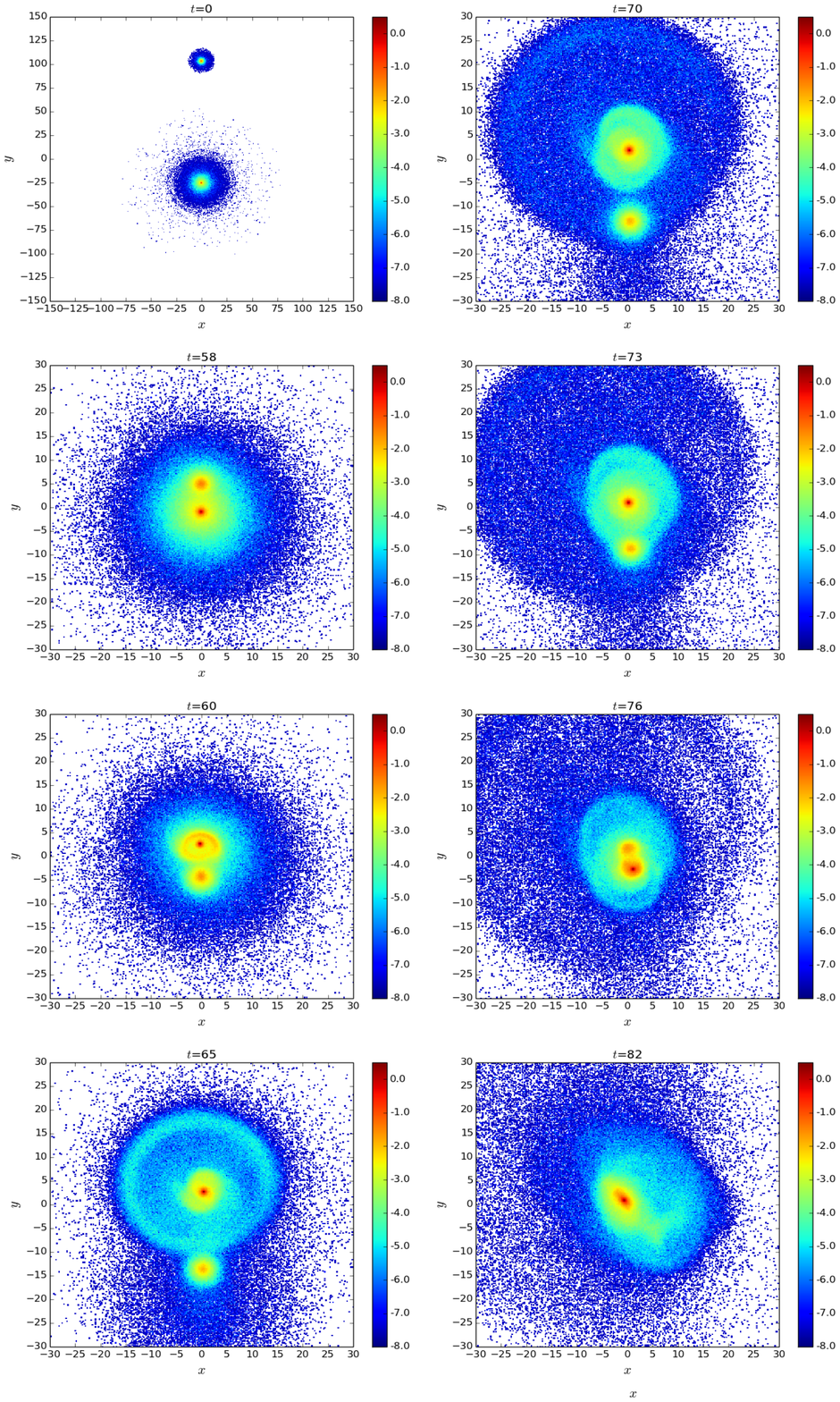}
 \caption[Time evolution of the projected density of the stellar components on the $x-y$ plane in Si40.]
{The evolution of the projected density of the stellar components, 
 including the stellar disk, the bulge and the dwarf's stellar part, 
 on the $x-y$ plane in Si40.
 The unit of length is kpc.
 The unit of surface density is $10^{10}M_{\odot}$ kpc$^{-2}$, and the scale
 is logarithmic.
 }
 \label {fig3:i00s4_xy_2D}
 \end{center}
 \end{figure} 
 
\clearpage
 \begin{figure}
 \begin{center}
 \includegraphics[trim=0cm 1.8cm 0cm 1.5cm,clip=true,angle=0,scale=0.7]{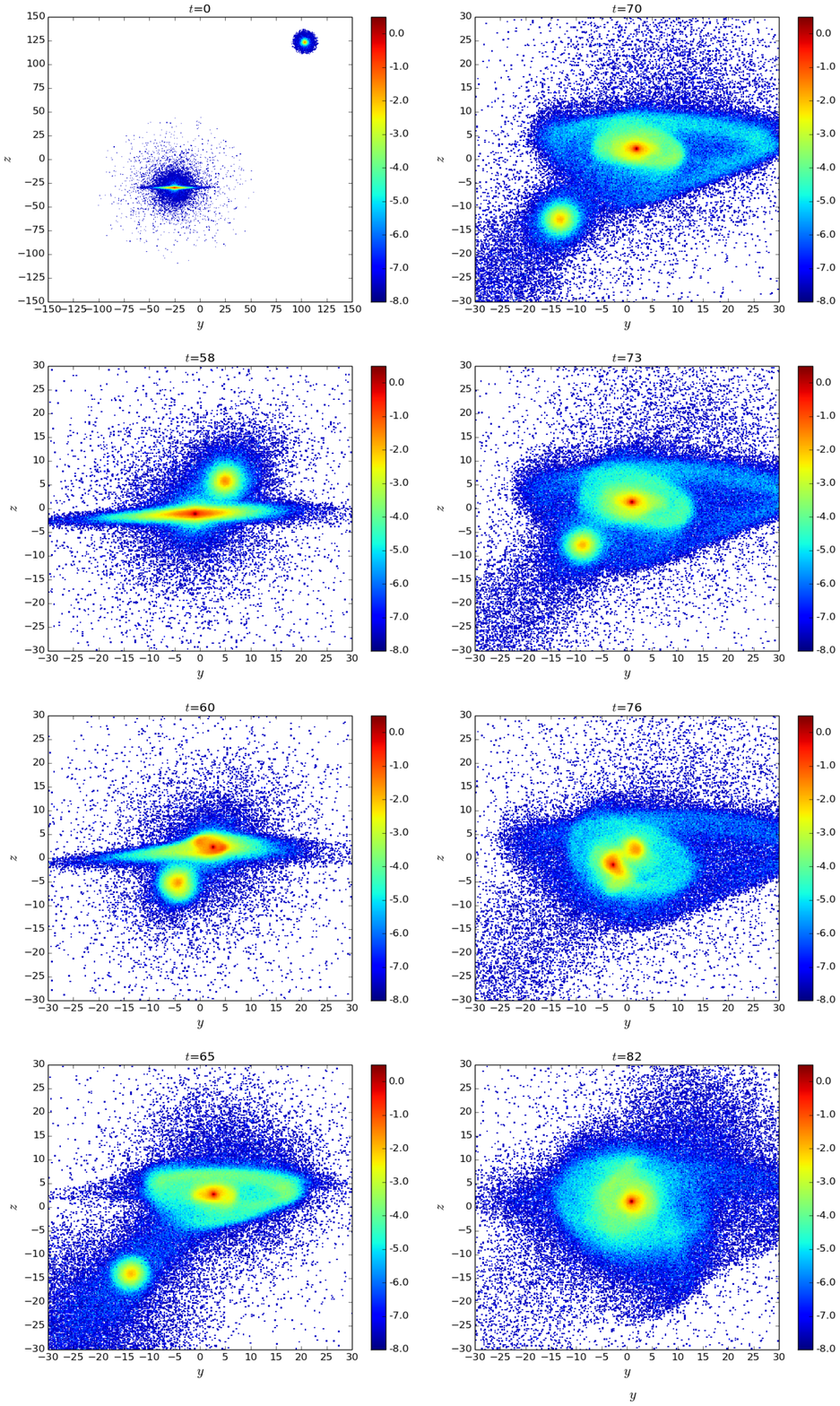}
 \caption[Time evolution of the projected density of the stellar components on the $y-z$ plane in Si40.]
{The evolution of the projected density of the stellar components, 
 including the stellar disk, the bulge and the dwarf's stellar part, 
 on the $y-z$ plane in Si40.
 The unit of length is kpc.
 The unit of surface density is $10^{10}M_{\odot}$ kpc$^{-2}$, and the scale
 is logarithmic.
 }
 \label {fig3:i00s4_yz_2D}
 \end{center}
 \end{figure}   
\clearpage
\begin{figure}
 \begin{center}
 \includegraphics[trim=1.5cm 1.2cm 0cm 1.5cm,clip=true,angle=0,scale=0.7]{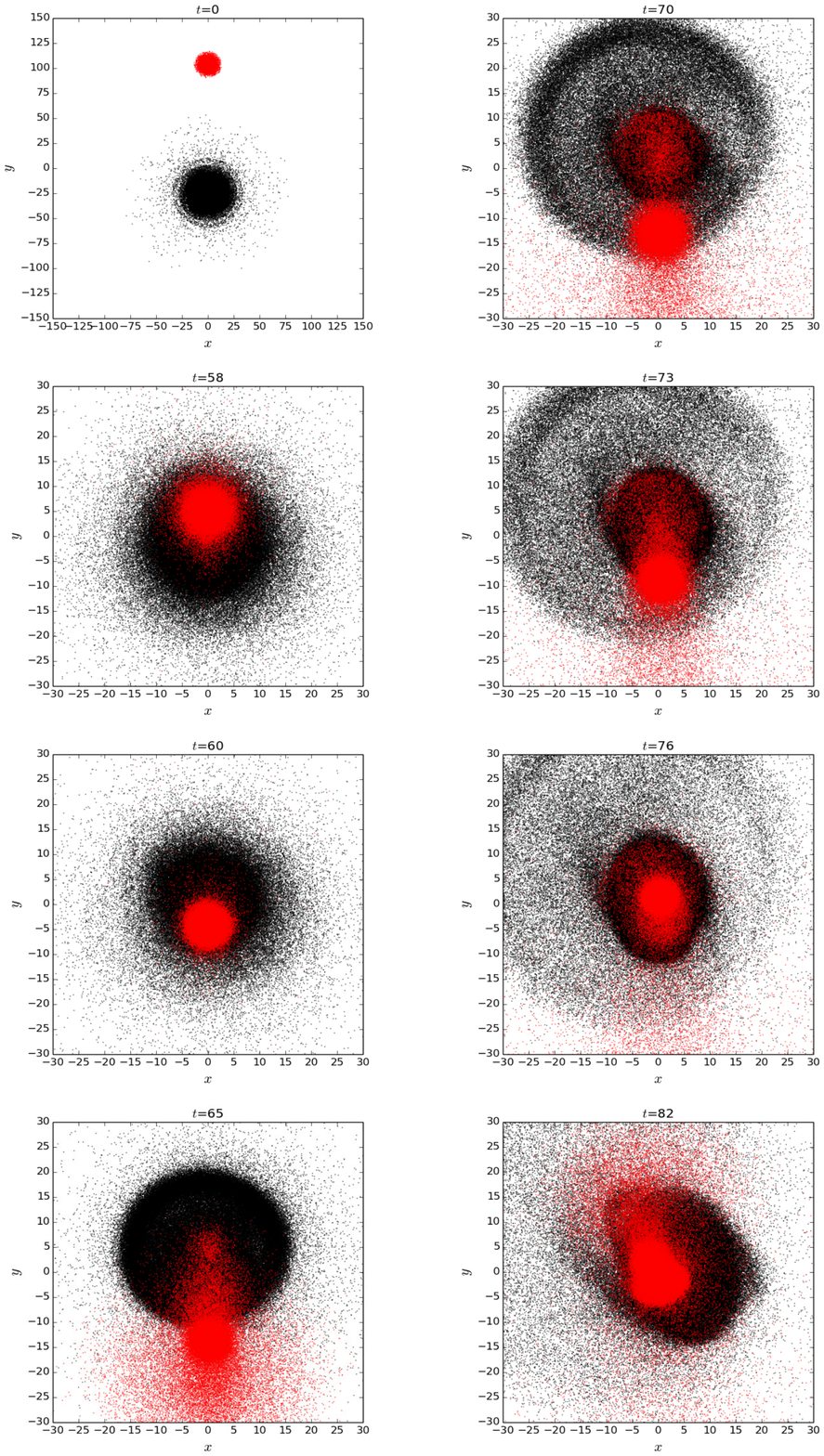}
 \caption[Time evolution of the stellar particles on the $x-y$ plane in Si40.]
{Time evolution of the stellar particles on the $x-y$ plane in Si40. 
 The black dots represent the stellar disk and the bulge particles 
of the disk galaxy.
 The red dots represent the dwarf's stellar particles. 
 The unit of length is kpc.
}
\label {fig3:com_xy_scatter}
 \end{center}
 \end{figure}
\clearpage
\clearpage
\begin{figure}
 \begin{center}
 \includegraphics[trim=1.5cm 1.2cm 0cm 1.5cm,clip=true,angle=0,scale=0.7]{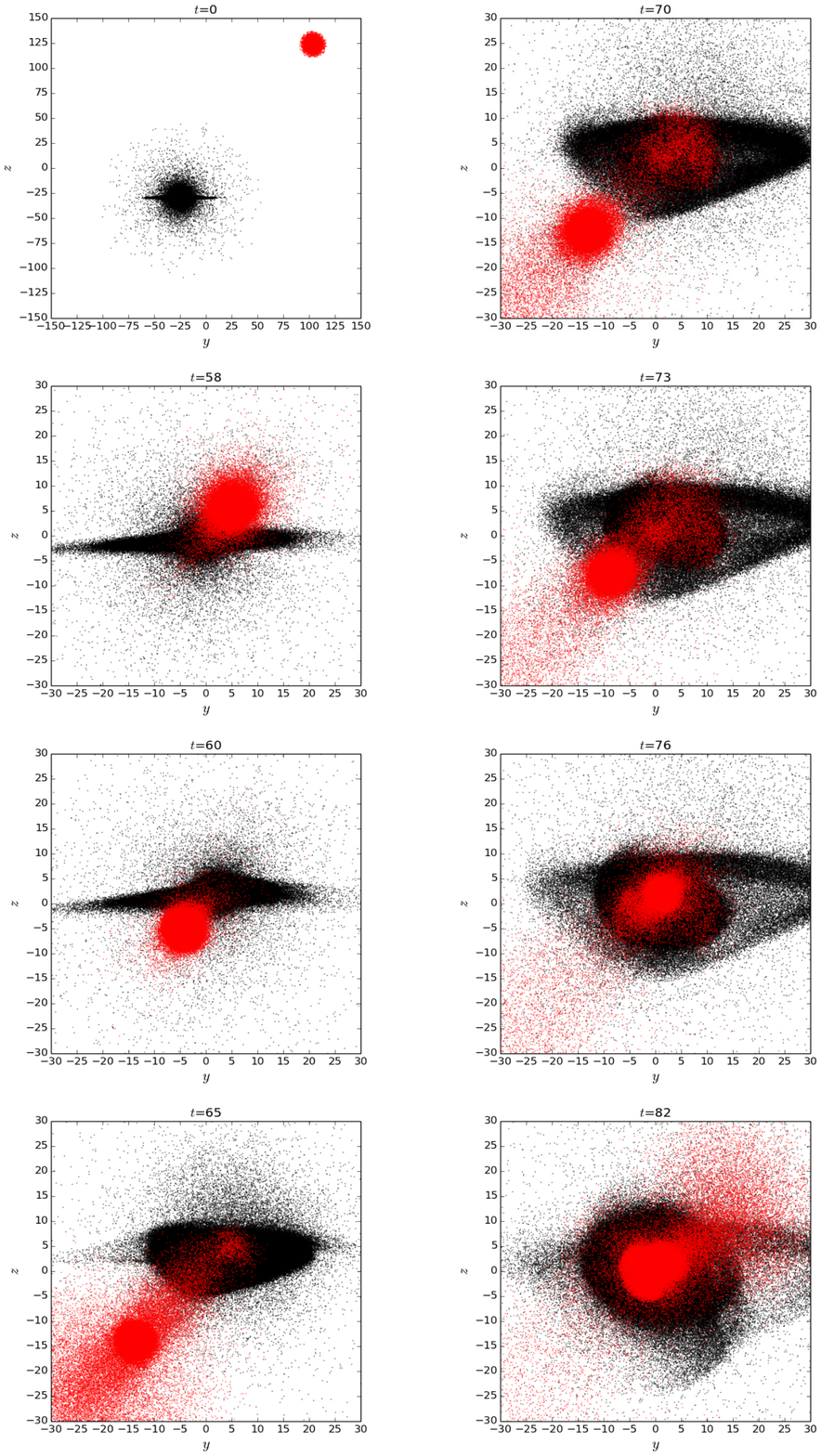}
 \caption[Time evolution of the stellar particles on the $y-z$ plane in Si40.]
{Time evolution of the stellar particles on the $y-z$ plane in Si40. 
 The black dots represent the stellar disk and the bulge particles
of the disk galaxy.
 The red dots represent the dwarf's stellar particles. 
 The unit of length is kpc.
}
\label {fig3:com_yz_scatter}
 \end{center}
 \end{figure}
\clearpage
 \begin{figure}
 \begin{center}
 \includegraphics[trim=1.5cm 7cm 0cm 6cm,clip=true,angle=0,scale=0.8]{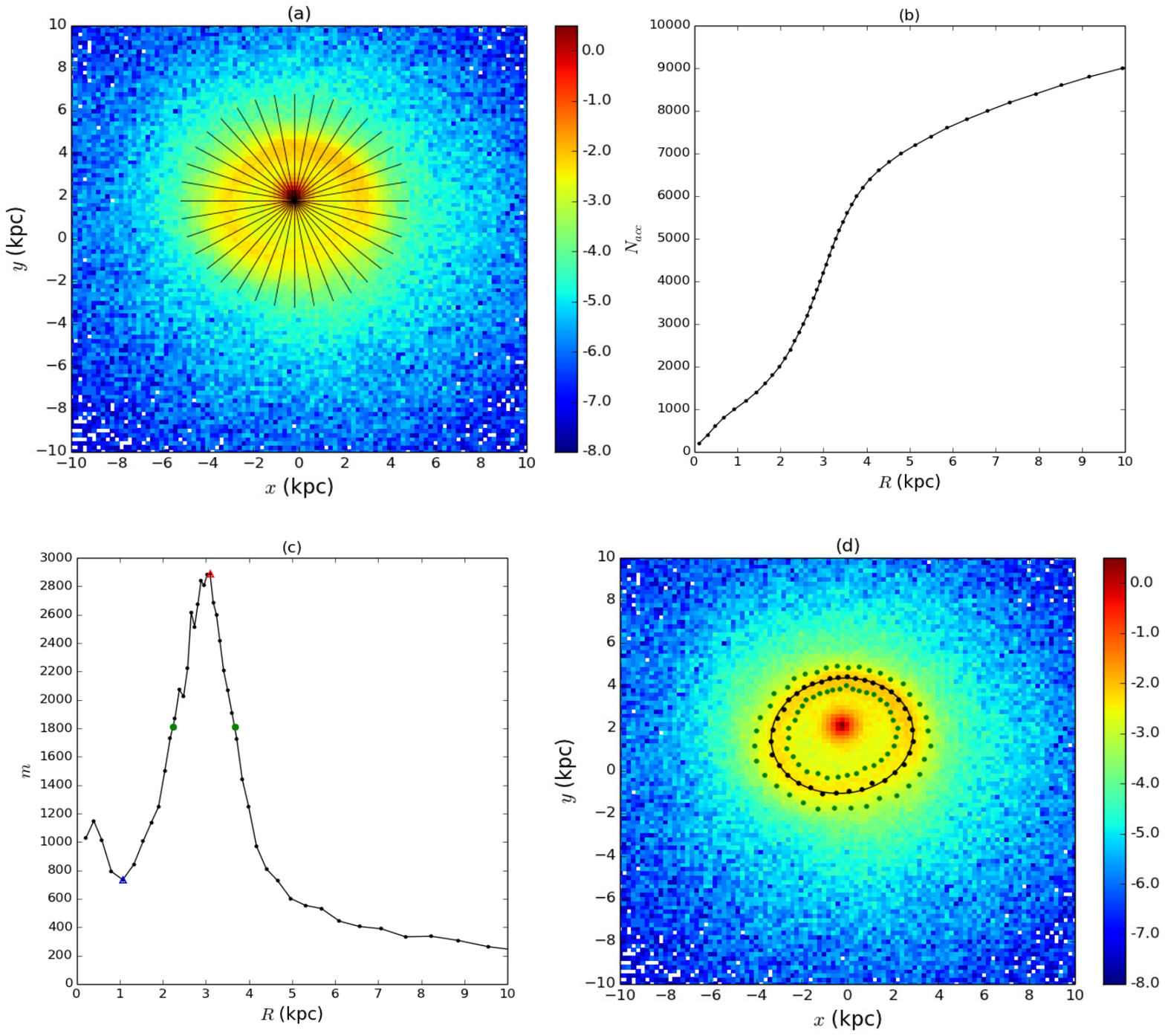}
 \caption[An example for the ellipse fitting process.]
{An example for the ellipse fitting process.
(a) The projected density of the stellar components of the disk galaxy, 
 including the stellar disk and the bulge, on the $x-y$ plane at $t=60T_s$ in simulation Si30. The azimuthal cells are also shown by 
those black grid-lines.
(b) The number of accumulated particles, $N_{acc}$, contained 
in one azimuthal cell as a function of radius $R$. 
(c) The slope, $m$, of the curve $N_{acc}(R)$, as a function of radius $R$. 
     The red triangle represents the maximum of the slope, and
     the blue triangle labels the minimum of the slope between the disk center 
     and the ring region.
     The green dots represent the slope which is equal to the average of 
     the maximum slope and the minimum slope, i.e. $m_{ave}$.  
 (d) The fitting ellipse (black curve) overlaid on the projected density. 
     The green dots represent the locations, $ri$ and $ro$, where 
     the slopes of $N_{acc}(R)$ are equal to $m_{ave}$ 
     for all different azimuthal cells. 
     The black dots represent the locations which are given by $(ri+ro)/2$ 
     for all different azimuthal cells.
 }
 \label {fig3:fitring}
 \end{center}
 \end{figure}
\clearpage
 \begin{figure}
 \begin{center}
 \includegraphics[trim=0cm 0cm 0cm 0cm,clip=true,angle=0,scale=0.9]
{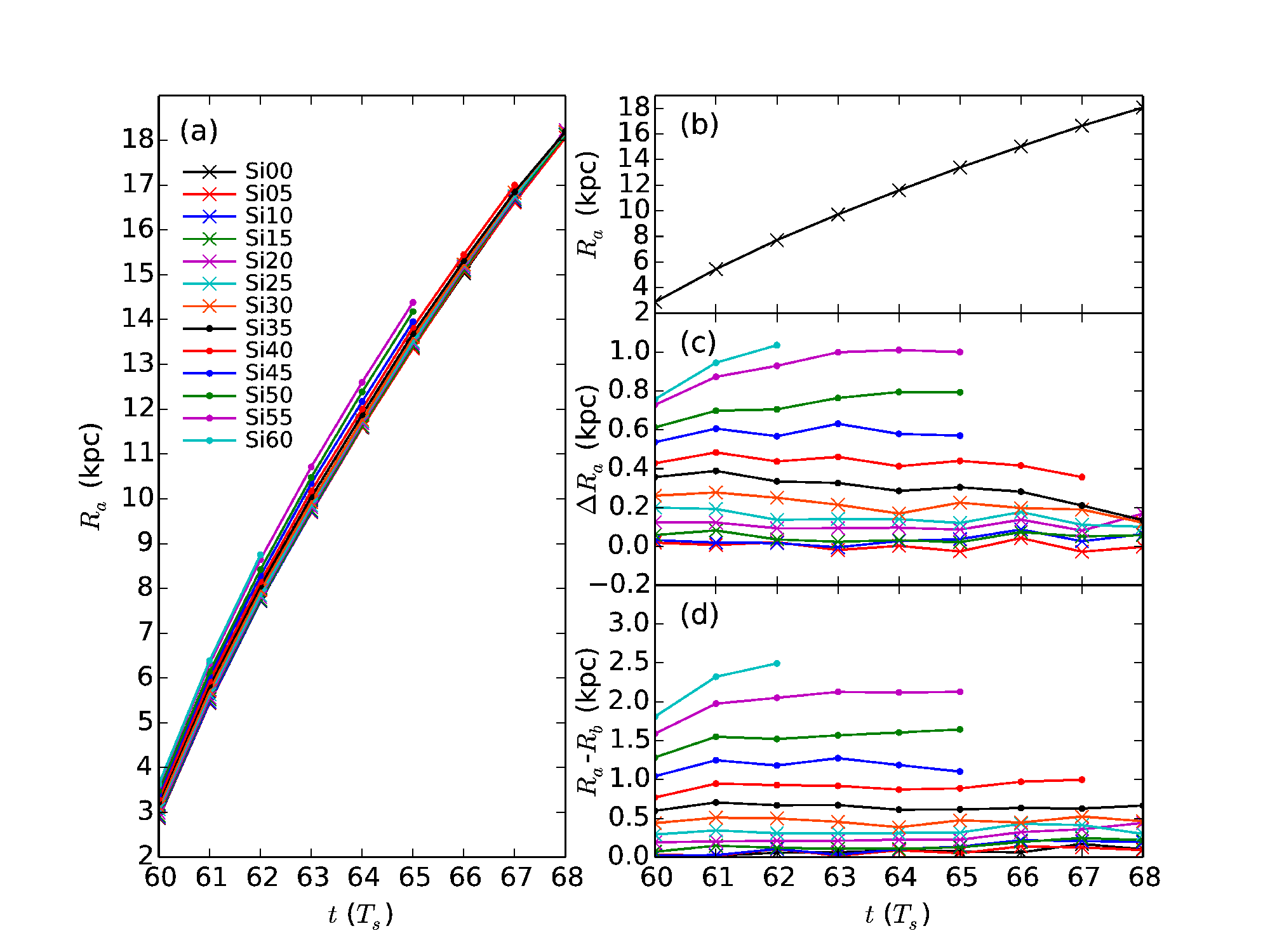}
 \caption[Time evolution of the location of rings in thirteen simulations.]
{The evolution of the sizes of rings in simulations. 
 Different lines represent different simulations, as indicated in panel (a).
 (a) The semi-major axis of the ring, $R_a$, as a function of time 
for different simulations.
 (b) The semi-major axis of the ring as a function of time for simulation Si00.
 (c) The differences of the semi-major axis between other simulations and Si00 
     as functions of time.
 (d) The differences between the semi-major axis and the semi-minor axis 
     as functions of time for different simulations.
 }
 \label {fig3:com_an}
 \end{center}
 \end{figure}
\clearpage
 \begin{figure}
 \begin{center}
 \includegraphics[trim=0cm 0cm 0cm 0cm,clip=true,angle=0,scale=0.9]
{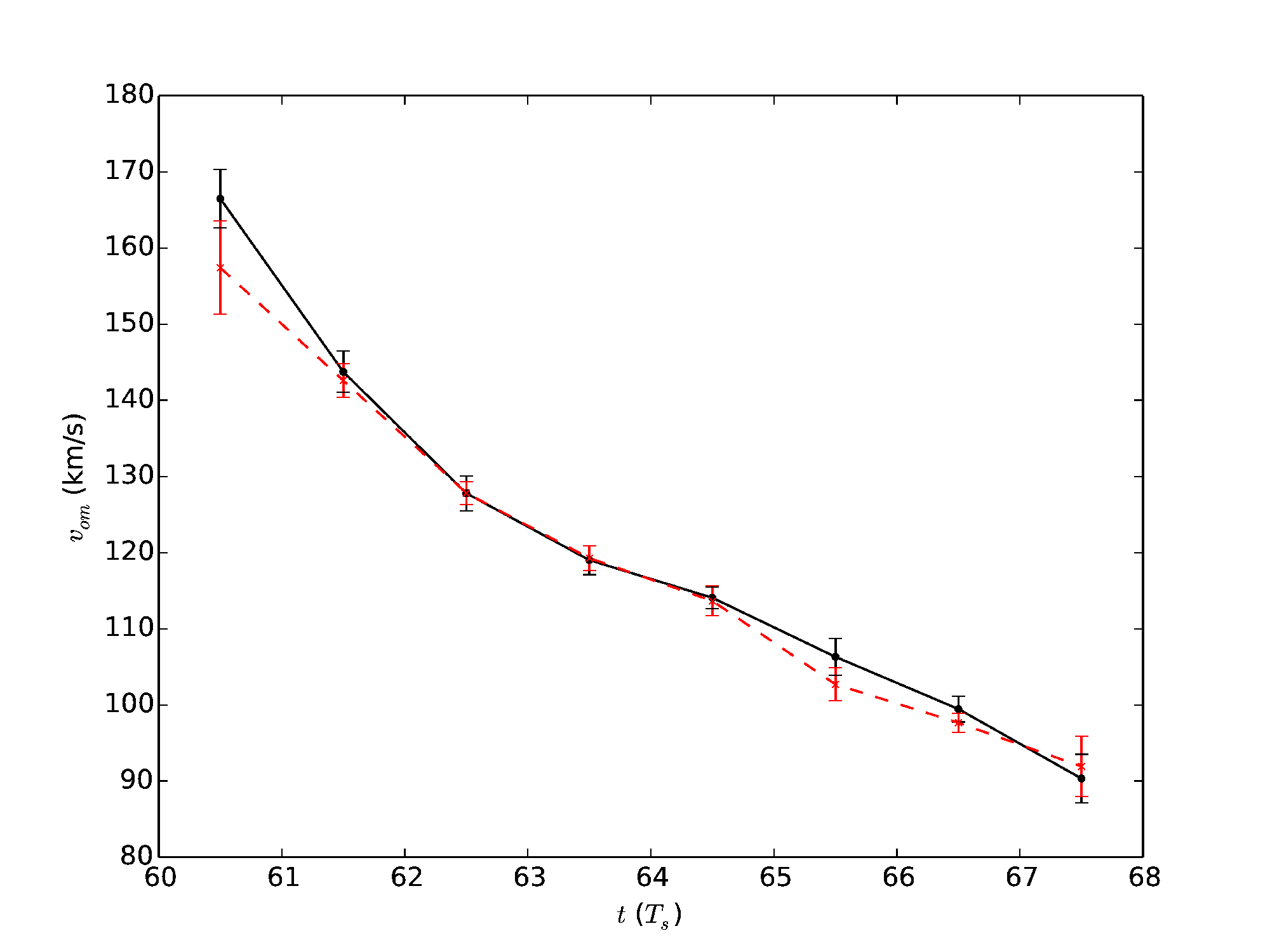}
 \caption[Time evolution of the out-moving speed of rings in thirteen simulations.]
{The evolution of the out-moving speeds, $v_{om}$, of rings in simulations. 
 The black line represents the average of the out-moving speeds of 
the semi-major axis for 13 simulations.
 The red line is for the semi-minor axis.
 The error bar indicates one standard deviation.
 }
 \label {fig3:com_vel}
 \end{center}
 \end{figure}
\clearpage
 \begin{figure}
 \begin{center}
 \includegraphics[trim=0.5cm 0cm 0cm 0cm,clip=true,angle=0,scale=1.0]
{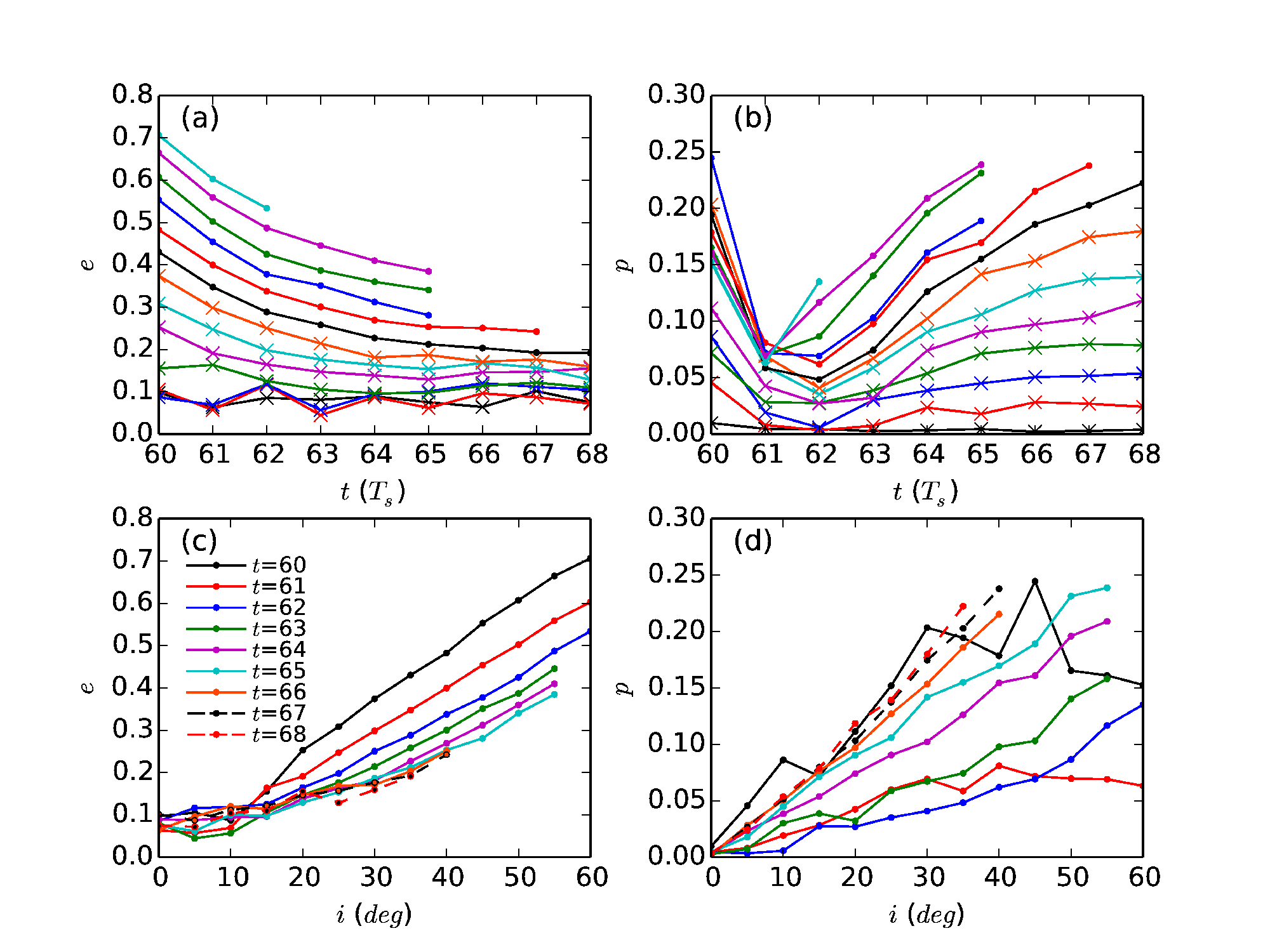}
 \caption[The properties of the rings in thirteen simulations.]
{The properties of rings in simulations.
 For panel (a) and (b), different lines are for different simulations 
as indicated in Fig. \ref{fig3:com_an}.
 (a) The eccentricity as a function of time. 
 (b) The offset $p$ as a function of time.
 For panel (c) and (d), different lines represent different simulation time, as indicated in panel (c).
 (c) The eccentricity as a function of the inclination angle, $i$.
 (d) The offset $p$ as a function of the inclination angle, $i$.
 }
 \label {fig3:com_an_2}
 \end{center}
 \end{figure}

\end{document}